\documentclass[11pt]{article}
\usepackage{amsmath}
\usepackage{latexsym}

\def\[{\left\lbrack}
\def\]{\right\rbrack}

\def\({\left(}
\def\){\right)}

\newcommand{\be}{\begin{equation}}
\newcommand{\ee}{\end{equation}}
\newcommand{\ea}{\end{eqnarray}}
\newcommand{\ba}{\begin{eqnarray}}

\begin{document}

\title{Fractional Canonical Quantization:\\   a Parallel  with Noncommutativity}
\author{Cresus F. L. Godinho$^{a}$\footnote{email:crgodinho@ufrrj.br},Jose Weberszpil$^{b}$,J. A. Helay\"{e}l Neto$^{c}$}
\maketitle

\begin{center}
{$^a$Grupo de F\' isica Te\'orica, 
Departamento de F\'{\i}sica, Universidade Federal Rural do Rio de Janeiro,\\
BR 465-07, 23890-971, Serop\'edica, RJ, Brazil\\
$^b$Universidade Federal Rural do Rio de Janeiro, UFRRJ-IM/DTL,\\
Av. Governador Roberto Silveira s/n , Nova Iguac\'{u}, RJ, Brazil\\
$^c$Centro Brasileiro de Pesquisas F\' isicas (CBPF),
Rua Dr. Xavier Sigaud 150, Urca, 22290-180, Rio de Janeiro, Brazil \\
}\end{center}


\begin{abstract}
\noindent Adopting a particular approach to fractional calculus, this paper sets out to build up a consistent extension of the Faddeev-Jackiw (or Symplectic) algorithm to carry out the quantization procedure of coarse-grained models in the standard canonical way.  In our treatment, we shall work with the Modified Riemman Liouville (MRL) approach for fractional derivatives, where the chain rule is as efficient as it is in the standard differential calculus.  We still present a case where we consider the situation of charged particles moving on a  plane with velo\-city $\dot{r}$, subject to an external and intense magnetic field in a coarse-grained scenario.  We propose an interesting parallelism with the noncommutative case. 
\end{abstract}

\maketitle

\pagestyle{myheadings}

\section{Introduction}


A powerful tool idealized in the sixties, motivating a large production of papers concerning constrained systems, the Dirac brackets (DB) \cite{3}, were an unmodified common point in the literature of the subject.  The main motivation for many works consisted in converting second-class systems into first-class ones.  The main objective was to obtain  gauge theories (first-class systems), the holy grail for the Standard Model.  Although not so popular as before, the analysis of constrained systems still draws a great deal of attention in the literature \cite{1}.

The Symplectic Method \cite{fj} is a geometrical way of approaching canonical quantization of constrained systems.  One of its main ingredients is the symplectic tensor,  that plays the role of a metric in the symplectic manifold.  Faddeev and Jackiw originally proposed a very interesting method which is centrally based on the attainment of the inverse of the symplectic tensor.  Its elements can be associated with the Dirac brackets and, consequently, it becomes possible to obtain the quantum commutators by means of the usual rule; of course, if no problems with the ordering of operators is present, then we can adopt the replacement below:
\begin{equation}
f^{-1}\,\rightarrow\,\{\,\,\,,\,\,\}^{\star}\rightarrow{\frac{1}{i \hbar}}\[\,\,\,,\,\,\]\,\,.
\end{equation}

The difficulty to obtain the symplectic tensor is proportional to the existence of constraints  \cite{BNW1,BNW2}.  It is important to highlight here that the constraints usually defined in Dirac's approach do not necessarily appear in the Symplectic Method.  The constrained systems considered in both scenarios have constraints usually referred to as true constraints.

It is well-known that several physical systems may be studied in a Lagrangian approach; their coordinates are usually embedded in a phase space.  Some of those Lagrangian systems may be written in a form where first-order time derivatives are present among the fields.  The so-called kinetic part is written in terms of first-order time derivatives which constitute 1-forms whose exterior derivatives appear in the equations of motion.  The resulting 2-form, the symplectic tensor, if singular,  probably signals a constrained system \cite{fj}. If the system is not constrained, the inverse of the symplectic tensor usually exists and provides the fundamental Poisson brackets.

Constrained systems can be usually solved by obtaining the inverse form of the symplectic tensor. Faddeev and Jackiw applied the Darboux theorem to work with canonical and non-canonical sectors separately \cite{BNW2}.  It is still possible to show that the Faddeev-Jackiw and the Dirac approaches are  completely equivalent whenever constrained systems are under consideration \cite{KS}.  However, it was shown that, if second-class constraints are present, the equivalence between the Dirac and Symplectic approaches fails \cite{Shirzad}.

On the other hand, there is a number of problems in considering classical systems besides the ones that involve the quantization of second-class systems,  as we have just mentioned above.  These problems encompass the so-called nonconservative systems.  Their peculiarity is that the great majority of actual classical systems is nonconservative but, in spite of that, the most advanced formalisms of classical mechanics deal only with conservative systems.
One way to suitably treat nonconservative systems it seems to be through Fractional Calculus, FC.  In a seminal paper, Riewe \cite{RW} had shown that was possible to build a kind of Lagrangian procedure to manage nonconservative systems.  However, more recently Dreisigmeyer and Young \cite{DY1,DY2} pointed out that Riewe's approach sometimes yields noncausal equations of motion.  They propose an alternative method to by-pass the problem of using fractional operators as kernels of Volterra series. Problems involving dissipation, for example, are present even at the microscopic level.  There is dissipation in every non-equilibrium or fluctuating process, including dissipative tunneling \cite{Cal} and electromagnetic cavity radiation \cite{Sen}, for instance.

Field-theoretic aspects of non-linear dynamics are today an important subject of study in different physical and mathematical sub-areas, but the real success and a radically new understanding of non-linear processes has occurred over the past 40 years. This understanding was inspired by the discovery and insight of chaotic dynamics, where the randomness of some physical processes are considered; more precisely, when particle trajectories are indistinguishable for random process \cite{GZ}.  


Fractional Calculus is one of the generalizations of the classical calculus.  It has been used in several branches of science. FC provides a redefinition of mathematical tools and it seems very useful to deal with anomalous and frictional systems.  For a very detailed review of many  advances and purposes, the readers should consider the references \cite{referee sug 1,JTM book} In particular, we can cite the continuous time random walk scheme as a physical counterpart example, where, within the fractional approach, it is possible to include external fields  in a straightforward manner. Also, the consideration of transport in the phase space spanned by both position and velocity coordinates is possible within the same approach. Moreover, the calculation of boundary value problems may be driven to a form analogous to the procedure adopted to treat the corresponding standard differential equations \cite{klafter,klafter2,klafter3,9,scalas,Hilfer2}.

Other important applications may be found by investigating response functions, for which many studies have been reported on the phenomenon of non-exponential, power-law relaxation which is typically observed in complex systems,  such as dielectric and ferroelectric systems.  Furthermore,  describing dynamical  processes in disordered or complex systems,  such as relaxation or dielectric behavior in polymers or photo bleaching recovery in biologic membranes,  has proven to be an extraordinarily successful tool.  The main feature of such systems is a strong (in general, randomic) interaction between their components in the passage to a state of equilibrium.  Some authors have proposed fractional relaxation models to describe filled polymer networks and investigate the dependence of a number of parameters on the filler content \cite{stanisvasky,Metzler}.  The study of exactly solvable fractional models of linear viscoelastic behavior is another successful field of application.  In recent years, both phenomenological and molecular-based theories for the study of viscoelastic materials have been proposed with integral or differential equations of fractional order. Some current models of viscoelasticity based on FC are usually derived from the Maxwell model by replacing the first-order derivative ($d/dt$) by its fractional version ($d^{\alpha}$/$dt^{\alpha}$) \cite{Glockle}, where $\alpha$ is not integer.  Presently, areas such as field theory and gravitational models demand new conceptions and approaches which might allow us to understand new systems and could help in extending well-known results. Interesting problems may be related to the quantization of field theories for which new approaches have been proposed  \cite{cresus,baleanu,goldfain}.    

In this work, we shall adopt FC to analyze the well-established canonical quantization symplectic algorithm.  The focus is to construct a generalized extension of that method  to treat a broader number of mechanical systems with respect to the standard method. In this sense, we shall adopt the Modified Riemann Liouville (MRL) prescription for fractional derivative.

Since FC has not yet actually been explored enough in field theory, despite some really interesting recent contributions \cite{cal1,goldfain,goldfain2,Herrmann,nos,nos2,referee sug 2}, we have tried to construct a self-sustained paper with content distributed as follows.  In Section (2), we provide  a short review on the symplectic algorithm along with its main equations and formulations.  Next, in Section (3), we present some basic aspects of the MRL approach that fit for our purposes here.  In Section (4), we propose the so-called fractional extension for the symplectic scheme on two different scenarios: the first one, in its simpler form, the Lagrangian has the following structure: $L=\eta^i f_{ij}\,{}_0D^{\alpha}_t\eta^j-V(\eta)$.  In the second, and more general scenario,  the Lagrangian considered is $L=a_i(\eta,\partial\eta)\,{}_0D^{\alpha}_t\eta^i-V(\eta)$, where in the kinetic sector, $(a_iD\eta)$, we now  have some general function of $\eta, \partial\eta$.  In both cases, by using the fractional time derivative ${}_0D^{\alpha}_t$ in the MRL formulation, we can  reassess the simplectic scheme and obtain new dynamical consequences, as well as  extensions for the simplectic matrix and Euler-Lagrange equations.  In Section (5), we present an interesting case study, where an electron charge is confined in a two-dimensional plane $(x,y)$ and subject to a constant magnetic field.  The coarse-grained equivalent system presents us some new interpretations and results, including some parallels with the classical and noncommutative quantum Hall effect.  Subsequently our Concluding Remarks are presented. We also present therein an example and comment on our results and possible new paths to be followed in forthcoming papers.

\section{Symplectic Quantization Scheme: Refreshment}
Following the symplectic algorithm \cite{fj,BNW1,BNW2}, the first-order formalism, one starts by considering a Lagrangian which is first-order in time derivative, represented with its kinetic and potential sectors, respectively
\begin{equation}
L^{(0)}\,=\,p_k\dot{q_k}-V(p,q),\,\,\,\, p_{k},q_{k},(k\,=\,1,2,\ldots,2n),
\end{equation}
where we adopt the summation convention for repeated indices and we consider it hereafter.
Introducing now $4n$ bosonic phase-space variables labeled here as $\eta$,  we can rewrite the Lagrangian in its canonical one-form
\begin{equation}
L^{(0)}\,=\,{1\over2} \eta_i f^{(0)} \dot{\eta}^i-V(\eta),
\end{equation}
where $f^{(0)}$ is called symplectic matrix
\begin{equation}
\label{ssm}
f^{(0)}\,=\,
\left[
\begin{array}{cc}
0 & 1 \\
-1 & 0 
\end{array}
\right].
\end{equation}
The equations of motion can be obtained by taking the variation $"\delta L\,=\,0"$,
\begin{equation}
\dot{\eta^{i}}\,=\,(f^{(0)})_{ij}^{-1}{{\partial} V\over{\partial \eta^{j}}}\,.
\end{equation}
Adopting now a more extensive viewpoint, let us write the Lagrangian
\begin{equation}
L^{(0)}\,=\,a_i(\eta)\dot{\eta}\,-\,V^{(0)}(\eta),\,\,\,\,(i=1,2,\ldots,m)
\end{equation}
It can be rewritten as
\begin{equation}
L^{(0)}dt\,=\,a_i(\eta)d{\eta}\,-\,V^{(0)}(\eta)dt,
\end{equation}
where, from the one-form $a_i(\eta)d{\eta}$, we can obtain the two-form
\begin{eqnarray}
f^{(0)}&=&d\[a_i(\eta)d{\eta}\] \nonumber \\
f^{(0)}&=&{1\over2}f^{(0)}_{i j}d\eta^{i}\wedge d\eta^{j} \nonumber \\
(i,j&=&1,\ldots,m),
\end{eqnarray}
and, consequently,
\begin{equation}
f^{(0)}_{i j}\,=\,{{\partial a_j}\over{\partial \eta^i}}\,-\,{{\partial a_i}\over{\partial \eta^j}}\,.
\end{equation}
It is well-known that the symplectic matrix in such cases is an antisymmetric tensor. From the symplectic tensor, it is possible to define the Euler-Lagrange equations 
\begin{equation}
\label{1}
\dot{\eta}^j\,=\,\[f^{(0)}_{i j}\]^{-1}{{\partial} V^{(0)}\over{\partial \eta^{i}}}\,.
\end{equation}
The main reason to approach the problem in this way is that the symplectic matrix has a very close relation to the Dirac brackets $\{\,\,,\,\,\}^{\star}$,
\begin{eqnarray}
f^{-1}_{ij}\,=\,\{\,\,,\,\,\}^{\star}.
\end{eqnarray}
A very important point should be pointed out when $f_{ij}$ is singular.  If this is the case, the symplectic tensor is not well-defined and the  bracket structures are not yet accessible.  This situation probably points to a constrained system and, as such, we must proceed in a non- standard way.  First of all,  $f_{ij}$ is not invertible and, according to the symplectic algorithm, we must obtain the zero-modes $\nu_n$ satisfying the following relation
\begin{eqnarray}
f^{(0)}_{i j}\nu^{(0)}_{nj}&=&0 ;
\end{eqnarray}
\\
combining it with (\ref{1}), we obtain the very useful relation
\begin{eqnarray}
\label{2}
f^{(0)}_{i j}\nu_n^{i}\dot{\eta}^j&=&\nu_n^{i}{{\partial} V^{(0)}\over{\partial \eta^{i}}}\,=\,0 .\nonumber \\
\end{eqnarray}
This may be a constraint, but not exactly in a Dirac-like sense.  These constraints are usually introduced in the Lagrangian by means of Lagrangian multipliers which "deform" the kinetic sector of the Lagrangian.  This can be done by taking the time-derivative of the constraint and making use of some Lagrange multiplier.  These procedures will enlarge the configuration space of the theory, so that we can identify new vectors,
\begin{eqnarray}
a_i^{(1)}&=&a_i^{(0)}\,+\,\lambda_m^{(0)}\partial_i \Omega_m^{(0)}
\end{eqnarray}
where $\Omega_m^{(0)}$ are the constraints obtained from (\ref{2}). However, it may also occur that we arrive at a point where we still obtain a singular matrix and the corresponding zero modes do not show us any new constraint.  This is a strict case of gauge theory.  In such a situation, it is necessary to impose some specific gauge condition. 

\section{Fractional Approach}

It is well-known that several definitions of fractional derivative and fractional integral exist;  for instance, Grunwald-Letnikov, Riemann-Liouville,
Caputo, Weyl, Feller, Erdelyi-Kober and Riesz fractional derivatives as well as fractional Liouville operators, which have been popularized when fractional integration is performed in the framework of the dynamical systems under study. Following this idea, let us consider an approach recently proposed which subtly modifies the usual Riemann-Liouville, known as MRL \cite{gj1,ric}.  Some authors believe that the Riemann-Liouville fractional derivative offers some disadvantages when certain real-world practical problems are being studied.  This could indicate future problems when some physical approach was necessary to understand some new phenomena based on fractional calculus \cite{Kai}.  In that sense is plausible to discuss a modified concept presented by Caputo.

The main reason to by-pass the piftails of Riemann-Liouville and Caputo definitions, is that, in the first definition, the derivative of a constant is not zero, and in the second one, it is necessary to have a higher-order derivative to evaluate the lower-order derivative.  Moreover, the chain rule, when considered in such cases, shows itself an un-practical exercise; but, in the MRL framework, it has almost the same shape as in the usual calculus, and for this particular reason it is more indicated for our purposes here.

In this sense, the fractional derivative of order $\alpha, \alpha < 0$ of a given function $f(x)$ is given by,
\begin{eqnarray}
\label{3}
f^{\alpha}(x)&=&{{1}\over{\Gamma(-\alpha)}} \int^{x}_{0} (x\,-\,\xi)^{-\alpha-1}\Delta f \,d\xi,\, \alpha < 0,  \nonumber \\
\big(f^{(\alpha-1)}(x)\big)^{\prime}&=&{{1}\over{\Gamma(1-\alpha)}}{{d}\over{dx}}\int^{x}_{0} (x-\xi)^{-\alpha}\Delta f \,d\xi \nonumber,\, 
\\
&0<& \alpha <1.
\end{eqnarray} 
where we have considered that
\begin{eqnarray}
\Delta f\,=\,f(\xi)-f(0).
\end{eqnarray}
Following this approach, let us consider now two functions, $f(x)$ and $u(x)$. Depending upon the nature of the functions, we can work with the following chain rules:
if $f(u)$ is $\alpha$-th differentiable and $u(x)$ is differentiable w.r.t. $x$ then
\begin{eqnarray}
{{d^{\alpha}f(u(x))}\over{dx^{\alpha}}}&=&{{d^{\alpha}f(u)}\over{du^{\alpha}}}\Big({{du}\over{dx}}\Big)^{\alpha},
\end{eqnarray} 
if $f(u)$ is differentiable w.r.t. $u$, but not differentiable w.r.t. $x$ and $u$ is $\alpha$-th differentiable w.r.t. $x$
\begin{eqnarray}
{{d^{\alpha}f(u(x))}\over{dx^{\alpha}}}&=&{{df(u)}\over{du}}\Big({{d^{\alpha}u}\over{dx^{\alpha}}}\Big).
\end{eqnarray}
An interesting observation here is that the MRL approach is near to Caputo's definition, but however, without the requirement that the fractionally derived function be also integer derivable as highlighted in \cite{Go1,Go2}.
Despite some criticism about this approach, it is worthy, in order to better understand and check
the whole formalism, to work possible different scenarios. More recently, a formal demonstration of the fractional
Taylor expansion, in the context of the MRL, has been worked out and is presented in \cite{lv}.
This is an important step for our present contribution, once our chain rule relies upon this fractional Taylor expansion.
We would like to remark here that we are not making any sort of criticism  to other approaches to fractional derivative; we have to remember that, today,  we can work with several different definitions. We really have a very rich set of options, but it seems that each definition possesses particular characteristics and, so, we must suitably apply them in some physical context.  In our case, we believe that the MRL set-up is sufficient to our main purposes, mainly by virtue of  its very simplified Leibniz and chain rules.  On the top of that, we are able to deal with non-differentiable functions in a coarse-grained context.

\section{Fractional Symplectic Approach}
In this section, we give an overview of how to obtain the symplectic algorithm through that alternative approach.  While we make no claims to originality, our point of view is very singular if compared with others accounts in the literature. Then let us consider a simplest case of some field theory where the action is given by
\begin{eqnarray}
S&=&\int d^4 x \left[\pi_k(x){}_0D^{\alpha}_t \phi_k(x)\,-\,H(\pi,\phi) \right]\nonumber \\
k&=&1,2,\cdots,2n
\end{eqnarray}
where now for our purposes we adopt ${}_0D^{\alpha}_t$ as time fractional derivative in MRL sense (\ref{3}).

We can rewrite the Lagrangian as
\begin{eqnarray}
S&=&\int d^4 x \left[{1\over2}\pi_k(x){}_0D^{\alpha}_t\phi_k(x)\,-\,{1\over2}\phi_k(x)_0D^{\alpha}_t\pi_k(x)\,-\,H\right]\,,\nonumber \\
\end{eqnarray}
it should be noted two points here: we have changed the notation for time derivative operator to avoid confusing with the main definition (\ref{3}), and the fields are not being considered in a coarse-grained context yet.  We are proposing only the fractional extension for action considering the fractional time derivative acting on the fields and its possible unfolds.   Introducing now $4n$ new (bosonic) phase space variables: $(\eta_i)\,=\,(p,q)\,,i=1,2,\cdots,4n\,\,$, it is possible to write the Lagrangian in a symplectic context,
\begin{eqnarray}
S\,=\,\left[{1\over2}\eta^i f^{(0)}\,\,{}_0D^{\alpha}_t\eta_i\,-\,H(\eta)\right],
\end{eqnarray}
where once more we find the same sympletic matrix given by (\ref{ssm}).
However a subtle and expected difference occurs when we obtain the equation of motion by the action variation, $\delta S\,=\,0$, implying a very reasonable relation
\begin{eqnarray}
{}_0D^{\alpha}_t\eta^i&=&[f^{(0)}]^{-1}_{ij}{{\partial H}\over{\partial \eta^j}}.
\end{eqnarray} 
The result above probably points to memory effect induced to a dynamical system when an
evolution described through a fractional differential equation is considered. By the general
definition (\ref{3}), the fractional derivative supposes such memory effects because of its dependence
on many time moments.  Its applicability could be justified for systems with dissipation effects where friction
forces would be present.  Very close result was obtained by \cite{Khalili} using Riemann-Liouville derivative. 

Realizing the fact that there are systems where is necessary to consider a more
detailed and deep Lagrangian description, it is interesting to think in a more general sense. 
In this way actions like
\begin{eqnarray}
\label{4a}
S&=&\int \left[a_i(\eta){}_0D^{\alpha}_t\eta^i\,-\,V(\eta)\right]d^4x,\,\,i\,=\,1,\cdots 4n
\end{eqnarray}
will be of great usefulness.    
It is well known that Lagrangians with higher order on time derivative can be put in the first-order formulation too, considering some specific and well tested field redefinitions. In order to become more explicit in our considerations about the canonical structure let us take once more the variation of action $\delta S\,=\,0$
\begin{eqnarray}
\delta S&=&\int d^4x \left[\delta a_i \,{}_0D^{\alpha}_t\eta^i+a_i \delta({}_0D^{\alpha}_t\eta^i)-\delta H\right] \nonumber \\
0&=&\int d^4x\left[{{\partial a_i}\over{\partial \eta^j}}\delta \eta^j {}_0D^{\alpha}_t \eta^i+a_i\,{}_0D^{\alpha}_t\delta \eta^i-\delta H\right] \nonumber \\
0&=&\int d^4x \left[{{\partial a_i}\over{\partial \eta^j}}\delta \eta^j {}_0D^{\alpha}_t \eta^i-{}_0D^{\alpha}_ta_i\delta \eta^i-{{\partial H}\over{\partial \eta ^j}}\delta \eta^j\right] \nonumber \\
0&=&\int d^4x \left[{{\partial a_i}\over{\partial \eta^j}}\delta \eta^j {}_0D^{\alpha}_t \eta^i-{{\partial a_i}\over{\partial \eta^j}}{}_0D^{\alpha}_t\eta^j \delta \eta^i-{{\partial H}\over{\partial \eta ^j}}\delta \eta^j\right] \nonumber \\
0&=&\int d^4x \left[\Big({{\partial a_i}\over{\partial \eta^j}}-{{\partial a_j}\over{\partial \eta^i}}\Big){}_0D^{\alpha}_t\eta^i-{{\partial H}\over{\partial \eta ^j}}\right]\delta \eta^j\,, \nonumber \\
\end{eqnarray}
if $\delta \eta^j$ are independent 
\begin{eqnarray}
{}_0D^{\alpha}_t\eta^i&=&[f^{(0)}_{ij}]^{-1}{{\partial H}\over{\partial \eta ^j}}.
\end{eqnarray}
where
\begin{eqnarray}
\label{symplectic matrix}
f_{ij}&=&{{\partial a_i}\over{\partial \eta^j}}-{{\partial a_j}\over{\partial \eta^i}},
\end{eqnarray}
is the usual simplectic matrix as considered before.

A very important point to be remarked here is that in spite our last result could seem 
usual, it is not, for reason that we obtained it at fractional context only because
we employed the chain rule in MRL context.  Nowadays, some interesting results may be found in the literature,
but always building theories and models through Riemann-Liouville or Caputo definitions; however,
in these cases in spite of the success of such approaches, the chain rule is always avoided.  

\section{Coarse-graining Embedding}
\subsection{Dirac-Bergmann Algorithm in the coarse-graining}
It has been seldom realized that some sort of coarse-graining is necessary in order that
typically quantum features of a system (with finite number of degrees of freedom) do not
dominate its appearance. The coarse-graining enters differently in different theories of quantum
to classical relation, and is not always equally strongly emphasized. In the theories of
decoherences \cite{Zu}, the emphasis is on the influence of the environment, but the description of
the environment must be coarse-grained to fulfill the desired decoherence effects.  In this way, we
believe that is important to adapt some quantizations procedures to this reality so we will consider now
a more generalized approach for symplectic algorithm.  We will start our attention understanding the behavior of
Dirac bracket on coarse-graining context and we will use again the MRL prescriptions for that purpose.  So, let us start considering some coarse-grained configu\-ration space, 
\begin{eqnarray}
\label{7}
{\Xi}^{\alpha}\,=\,\{ {\bf q}_i,{\bf p}_i,{\bf t}\}
\end{eqnarray}
we have not adopted the label $\{q^{\alpha}_i,p^{\alpha}_i,t\}$ in the coordinates to let a more simplified notation and avoid suffix saturation, so we understand now that bold face coordinates are embedded in some fractional space-time context.  Let us consider an action within that $\alpha-$phase space
\begin{eqnarray}
\label{4}
S\,=\,\int_{t}^{t^{\prime}}\left[{\bf p}_i \,{}_0D^{\alpha}{}_t{\bf q}_i\,-\,H(\bf p,q)\right](dt)^{\alpha}
\end{eqnarray}
where the summation convention for repeated suffixes was considered once more. The variation of (\ref{4}) implies
\begin{eqnarray}
\delta S&=& \int_{t}^{t^{\prime}} \bigl[\delta^{\alpha}{\bf  p}_i\, {}_0D^{\alpha}_t{\bf q}_i+{\bf p}_i {}_0D^{\alpha}_t\delta^{\alpha}{\bf q}_i +\nonumber \\
&-&{{\partial^{\alpha}H}\over{\partial {\bf p}^{\alpha}_i}}(\delta {p}_i)^{\alpha}-{{\partial^{\alpha}H}\over{\partial {\bf q}^{\alpha}_i}}(\delta {q}_i)^{\alpha}\bigl](dt)^{\alpha}\,=\,0\nonumber \\
\end{eqnarray}.

Based on MRL formulation  \cite{gj1}, we can employ the following approximation,
\begin{eqnarray}
(\delta u_i)^{\alpha}\,\approx\,(\alpha !)^{-1}\delta^{\alpha} {\bf u}_i\,,
\end{eqnarray}
and, after some little algebra, we can obtain the fractional extension for Hamilton-Jacobi equations of motion, within the MRL context  
\begin{eqnarray}
\label{5}
{}_0D^{\alpha}_t {\bf q}_i&\approx&(\alpha !)^{-1}{{{\partial^{\alpha}}H\over{{\partial}{\bf p}^{\alpha}_i}}}\nonumber \\
{}_0D^{\alpha}_t {\bf p}_i&\approx&-(\alpha !)^{-1}{{{\partial^{\alpha}}H\over{{\partial}{\bf q}^{\alpha}_i}}}.\nonumber \\
\end{eqnarray}
Considering now some dynamical variable embedded in some coarse-grained spacetime , $\Theta\,=\,\Theta({\bf q}_i,{\bf p}_i,{\bf t})$ for instance, it is reasonable to figure its expansion by means of fractional Taylor's series  \cite{gj1}
\begin{eqnarray}
_0D^{\alpha}_t\Theta&=&{{\partial^{\alpha} \Theta}\over{\partial {\bf q}^{\alpha}_i}}(Dq_i)^{\alpha}+{{\partial^{\alpha} \Theta}\over{\partial {\bf p}^{\alpha}_i}}(Dp_i)^{\alpha}+{{\partial^{\alpha} \Theta}\over{\partial t^{\alpha}}}(Dt)^{\alpha} \nonumber\\
_0D^{\alpha}_t\Theta&\approx&{1\over{\alpha!}}\Big({{\partial^{\alpha} \Theta}\over{\partial {\bf q}^{\alpha}_i}}({}_0D^{\alpha}_t{\bf q}_i)+{{\partial^{\alpha} \Theta}\over{\partial {\bf p}^{\alpha}_i}}({}_0D^{\alpha}_t{\bf p}_i)\Big)+{1\over{\alpha!}}{{\partial^{\alpha} \Theta}\over{\partial t^{\alpha}}}({}_0D^{\alpha}_t{\bf t}).\nonumber \\
\end{eqnarray}
Using the Hamilton-Jacobi equations (\ref{5}) we can adapt an approximated fractional version (MRL sense) for the Poisson bracket
\begin{eqnarray}
\{U,V\}_{\alpha}&=&\Big({1\over{\alpha!}}\Big)^2\Big({{\partial^{\alpha} U}\over{\partial {\bf q}^{\alpha}_i}}{{{\partial^{\alpha}}V\over{{\partial}{\bf p}^{\alpha}_i}}}-{{\partial^{\alpha} U}\over{\partial {\bf p}^{\alpha}_i}}{{{\partial^{\alpha}}V\over{{\partial}{\bf p}^{\alpha}_i}}}\Big),\nonumber \\
\end{eqnarray}
where $U$ and $V$ are two dynamical variables defined on phase space.
It must be considered here that, though we have presented an approximated version, it is very reasonable to treat the quantization of nonlinear systems.  It is clear that there exists an extensive literature descri\-bing techniques for relating Lagrangian physical systems in the linear regime; however, it was only over the past 15 years or so that there
has been a general awareness of the possibility that irregular-looking fluctuations may be caused by deterministic chaotic dynamics \cite{Cv} in this sense we believe that new perspectives for quantize these systems must be investigated and MRL symplectic approach can be useful to research new results.

In a more general sense it is possible to obtain the right canonical structure for some fractional constrained system. 
Roughly speaking, the natural extension for the Dirac bracket obey the same algebraic protocol, but it starts with some Hamiltonian,
\begin{eqnarray}
\tilde{H}\,=\,H\,+\,\lambda_m \phi_m\,,
\end{eqnarray}
where $H$ is the canonical Hamiltonian, $\lambda_m$ are Lagrange multiplyiers and $\phi_m$ are coarse-grained constraints, more precisely
\begin{eqnarray}
\phi\,=\,\phi({\bf q}_i,{\bf p}_i)\,.
\end{eqnarray}
Following the usual steps, we can obtain Hamilton-Jacobi equations for constrained systems in a coarse-grained scenario,
\begin{eqnarray}
\label{6}
_0D^{\alpha}_t{\bf q}_i&\approx&(\alpha !)^{-1}\Big({{{\partial^{\alpha}}H\over{{\partial}{\bf p}^{\alpha}_i}}}+\lambda_m{{\partial^{\alpha} \phi_m}\over{\partial}{\bf p}^{\alpha}_i}\Big)\nonumber \\
_0D^{\alpha}_t{\bf p}_i&\approx&-(\alpha !)^{-1}\Big({{{\partial^{\alpha}}H\over{{\partial}{\bf q}^{\alpha}_i}}}+\lambda_m{{\partial^{\alpha} \phi_m}\over{\partial}{\bf q}^{\alpha}_i}\Big),\nonumber \\
\end{eqnarray}
and, consequently, we can write the right form for the coarse-grained Dirac bracket
\begin{eqnarray}
\{U,V\}^{\star}_{\alpha}\,=\,\{U,V\}_{\alpha}-\{U,\phi_i\}_{\alpha}\[C^{\alpha}_{ij}\]^{-1}\{\phi_j,V\}_{\alpha}\,, \nonumber \\
\end{eqnarray}  
where $C^{\alpha}_{ij}$ is the matrix of constraints, usually defined on Dirac algorithm, in our case we will define it subsequently in (\ref{Dirac Matrix}). 

\subsection{Linking Symplectic and Dirac-Bergmann algorithms in the coarse-grained Scenario}
Our goal now shall be to link both algorithms through MRL \footnote{once more it is important to emphasize here that $\alpha$ suffix is the fractional degree of our coarse-graining physical space, no confusion must be done with Lorentz covariant suffix or other ones.} prescription.  So, let us first consider an action embedded in some coarse-grained space-time, 
\begin{eqnarray}
S&=&\int (d x)^{\alpha} \left[a_i({\boldsymbol \eta}){}_0D^{\alpha}_t{\boldsymbol \eta}_i(x)\,-\,V(\eta)\right]\,,\,\,k=1,2,\cdots,2n\,.\nonumber \\
\end{eqnarray}
Once more, we are dealing with a more simplified notation where the field variables does not have any fractional tag, therefore we adopt the same prescription and use bold face variables, so the action above gives rise to the Euler-Lagrange equations for our fractional scenario,
\begin{eqnarray}
f^{\alpha}_{ij}{}\,_0D^{\alpha}_t{\boldsymbol{\eta}}^i\,=\,{{\partial^{\alpha}H}\over{\partial {\boldsymbol \eta}^{\alpha j}}},
\end{eqnarray}
where the $f^{\alpha}_{ij}$ is the symplectic matrix. 

The so-called primary constraint is then
\begin{eqnarray}
\Omega_i\,=\,p_i-a_i\approx 0\,.
\end{eqnarray}
The Dirac bracket structure defines a matrix of constraints $C_{ij}$, in our case we have
\begin{eqnarray}
\label{Dirac Matrix}
C^{\alpha}_{ij}\,=\,\{\Omega_i,\Omega_j\}_{\alpha}\,=\,(\alpha!)^{-2}\Big({{\partial^{\alpha} a_i}\over{\partial {\bf \eta}^{\alpha j}}}-{{\partial^{\alpha} a_j}\over{\partial {\bf \eta}^{\alpha i}}}\Big),
\end{eqnarray}
or in a more concise way
\begin{eqnarray}
\label{7}
C^{\alpha}_{ij}\,=\,(\alpha!)^{-2}f^{\alpha}_{ij}\,.
\end{eqnarray}
Considering (\ref{7}) we can establish the right connection between both approaches
\begin{eqnarray}
\{\eta_i\,,\,\eta_j\}^{\star}_{\alpha}&=&\{\eta_i\,,\,\eta_j\}_{\alpha}-\{\eta_i\,,\,\phi_r\}_{\alpha}(\alpha!)^2[f^{\alpha}_{rs}]^{-1}\{\phi_s,\eta_j\}_{\alpha}\nonumber \\
\{\eta_i\,,\,\eta_j\}^{\star}_{\alpha}&=&(\alpha!)^{-2}\Big[(\delta_{ir})[f^{\alpha}_{rs}]^{-1}(\delta_{js})\Big] \nonumber \\
\{\eta_i\,,\,\eta_j\}^{\star}_{\alpha}&=&(\alpha!)^{-2}[f^{\alpha}_{ij}]^{-1}\,.
\end{eqnarray}
Of course that the same steps considered in the section II can be repeated here.  We obtained a new corrected symplectic matrix and reviewing the chain of reasoning it becomes apparent that we extended a well known quantization method to study the response of a coarse-grained based system when the transition from classical to quantum is considered.

\section{Physical Example}
\subsection{A Little Bit on the Quantum Hall Effect}
It is well-known that Integer Quantum Hall Effect may be understood in terms of the properties of independent electrons. 
Even though it is a well-tested result, the usual case was considered in \cite{Jackiw}.  The orbital dynamics of charged particles, more precisely electrons, moving on a  plane with velocity $\dot{r}$, subject to an external magnetic field, modulus $B$, perpendicular to the plane, has the following  action 
\begin{eqnarray}
S\,=\,\int_{r_1}^{r_2}\Big({1\over2}m\dot{r_i}^2\,-\,eA^i\dot{r_i}\Big).
\end{eqnarray}
Its respective equation of motion, governed by the Lorentz force in the Landau gauge is given by
\begin{eqnarray}
\ddot{r}^j={{eB}\over m}\epsilon^{ij}\dot{r}_{i},
\end{eqnarray}
Our goal now is to implement the fractional approach presented in Section III; in this sense, we shall apply the fractional temporal operator.  We must observe that this operator will impose a new time unity $(\sf{T}^{\alpha})$, so, for that reason, we will re-scale all units of length with the help of label  $(\sf{L}^{\alpha})$.  Physically, we could think of some heterogeneous planar system with different layers and different densities.  Today, graphene monolayers show that genuinely planar systems are physically realizable.  

For that purpose, the natural fractional extension is given by the following Lagrangian,
\begin{eqnarray}
\label{Hall}
L\,=\,k^{\alpha}\Big[{1\over2}m \,\,\,\Big({}_0D^{\alpha}_t r_i\Big)^2\,+\,e A^{i}\,\, {}_0D^{\alpha}_t r_i\,-\,V(r)\Big] \nonumber \\
\end{eqnarray}
We know that the properties of fractional calculus (derivatives and integrals) are not the same as in the usual calculus.  Therefore, we believe that with this approach, we can access some new perspectives, for instance, in complex systems where memory effects are very useful to describe suitably their dynamics.  The equation of motion for the Lagrangian (\ref{Hall}) is \footnote{it is important to remark that $D^{2\alpha}f\ne D^{\alpha}D^{\alpha}f$ and\\$k^{\alpha}$ is a constant to adjust the Lagrangian unit for Joule. It will remains equal $1$ for $D^{\alpha}\approx\Gamma(1+\alpha)D$} 
\begin{eqnarray}
{}_0D^{\alpha}_t{}_0D^{\alpha}_t\,r^{i}={2{e}\over m}\,\,{}_0D^{\alpha}_t\,A^{i},
\end{eqnarray}
and, by choosing the Landau symmetric gauge,
\begin{eqnarray}
A^{i}&=&{1\over2}B\epsilon^{ij}\,r_j
\end{eqnarray}
leaves
\begin{eqnarray}
\label{HallMRL}
{}_0D^{\alpha}_t{}_0D^{\alpha}_t\,r^{j}\,=\,{{eB}\over m}\epsilon^{ij}\,\,{}_0D^{\alpha}_t\,r_{i}\,.
\end{eqnarray}
If we consider the MRL differential relation $D^{\alpha}f\,\approx\,\Gamma(1+\alpha)Df$ once more, the differential equation (\ref{HallMRL}) can be cast under the form
\begin{eqnarray}
\label{HallMRL2}
\ddot{r}^j\approx{eB\over {m\Gamma(1+\alpha)}}\epsilon^{ij}\dot{r}_{i}\,.
\end{eqnarray}
If we look upon the Hall effect in the usual context, we know that the quantized theory yields the Landau levels when external forces are not involved. So, in that regime, and performing the change for the complex notation, considering $z=x+iy$, we rewrite the expression (\ref{HallMRL2}) as
\begin{eqnarray}
\ddot{z}\approx{i\omega\over{\Gamma(1+\alpha)}}\dot{z},
\end{eqnarray}
giving rise to us the following solution,
\begin{eqnarray}
z\approx z_0\,+\,d\,exp\Big({{i\omega\over{\Gamma(1+\alpha)}}t}\Big)\,,
\end{eqnarray}
where $z_0$ is usually considered arbitrary and called, guiding center, the constant $d$ is the radius o cyclotron.  
\subsection{Estimating the Fractionality}
We understand that there is an important and interesting discussion under consideration.  It is possible to notice that the cyclotron frequency gets a fractional correction $\omega^{\alpha}=\omega \big(\Gamma(1+\alpha)\big)^{-1}$, that our approach allows us to compute. In this context, we intend to investigate the two possible limits for $\alpha$, and we shall base our analisys by requiring that the ratio below is close to the relative error in the cyclotron frequency,
\begin{eqnarray}
\label{omegaalpha}
{{\omega_{\alpha}-\omega}\over{\omega}}&\approx&10^{-8}
\end{eqnarray}

In the first limit, we consider  $(\alpha\rightarrow0)$; then, we can take the gamma function after its usual expansion 
\begin{eqnarray}
\Gamma(\alpha+1)&=&\alpha\Gamma(\alpha)\,=\,1-\alpha\gamma, 
\end{eqnarray}
where $\gamma=0.57721$ is the Euler-Mascheroni constant.
Consequently, using the geometric expansion, the cyclotron frequency can be re-written now as
\begin{eqnarray}
\omega^{\alpha}={\omega\over{1\,-\,\alpha\gamma}}\,=\,\omega(1\,+\,\alpha\gamma).
\end{eqnarray} 

Using the relation (\ref{omegaalpha}),
we can write that
\begin{eqnarray}
\alpha&\approx&10^{-8}.
\end{eqnarray}

Curiously, $\alpha$ displays an interesting correlation with the noncommutative parameter $\theta$ \cite{Sus,Michael Douglas}; indeed the order of magnitude for the magnetic field is the same as the one in the Hall materials.

Now, let us consider the limit when $(\alpha\rightarrow1^-)$. We shall adopt a similar, but not quite the same, approach as the one in \cite{Tara}.  Let us apply once more the ratio (\ref{omegaalpha}) 
\begin{eqnarray}
{{\omega_{\alpha}-\omega}\over{\omega}}&\approx&10^{-8} \nonumber \\
{{1\over{\Gamma(1+\alpha)}}-1}&\approx&10^{-8} \nonumber \\
\alpha&\approx&0.9999999763473
\end{eqnarray} 
after considering the well-known experimental values and error bars of the Hall effect metrology \cite{jecke}.  It is important to remark that our last result points to the so-called low-level fractionality limit.

\subsection{Quantization and Connection with Noncommutativity}
Let us now consider the Lagrangian (\ref{Hall}) in the limit of intense $B$ and $(m\rightarrow0)$ as taken in the same gauge: 
\begin{eqnarray}
L\,=\,k^{\alpha}\Big[{1\over2}\,e B \epsilon^{ij}r_j\,\,\, {}_0D^{\alpha}_t r_i\,-\,V(r)\Big].
\end{eqnarray}
It is clear that we are dealing with a Lagrangian like (\ref{4a}), where the canonical pair is given here by 
\begin{eqnarray}
(eB\Gamma(1+\alpha)r_i\,,\,r_j), 
\end{eqnarray}
wich allows us to make the right identification $H\,=\,V(r)$.  Applying the canonical formalism, by means of symplectic algorithm, we can observe that our set of symplectic variables are $\eta={\{r_1\,,\,r_2\}}$ or ${\{x\,,\,y\}}$ and the kinetic parts read
\begin{eqnarray}
a^i_{r}\,=\,{1\over2}eB\Gamma(1+\alpha)\epsilon^{ij}r_j\,,
\end{eqnarray}
Following (\ref{symplectic matrix}), we obtain by direct calculation
\begin{eqnarray}
f^{ij}\,=\,eB\Gamma(1+\alpha)\epsilon^{ij},
\end{eqnarray} 
which yields the Dirac bracket
\begin{eqnarray}
(f^{ij})^{-1}\,=\{r_i\,,\,r_j\}^{\star}\,=\,\,\frac{\epsilon_{ij}}{eB\Gamma(1+\alpha)}.
\end{eqnarray}
The last equation can be re-written if we again consider the limit $(\alpha\rightarrow0)$. The Dirac bracket is 
\begin{eqnarray}
\{r_i\,,\,r_j\}^{\star}\,=\,\,{\frac{\epsilon_{ij}}{eB}}(1\,+\,\alpha\gamma),
\end{eqnarray} 
and the commutator
\begin{eqnarray}
\label{commutator}
\[r_i\,,\,r_j\]\,=\,i\hbar{\frac{\epsilon_{ij}}{eB}}(1\,+\,\alpha\gamma).
\end{eqnarray}
The form of the commutator can be understood as a contribution from the fractionality,  even if it is small.  The usual and expected noncommutative structure, given by 
\begin{eqnarray}
\[r_i\,,\,r_j\]\,=\,i\theta_{ij}
\end{eqnarray} 
gets a fractional correction in the limit $(\alpha\rightarrow0)$.  This interesting new perspective  may be putting in evidence a reasonable path for helping future investigations on high energy physics at Planck scale.  Another important point is that, in this limit, $\theta$ and $\alpha$ seems to have the same order of magnitude.

The noncommutativity of the space coordinates should not after all be surprising  in a coarse-grained framework.  There is always an intrinsic uncertainty in the position coordinates, so that a noncommutative character is very reasonable.  Moreover such an equivalence was mapped recently   
\cite{cal2}, where a connection between multifractional and noncommutative spacetimes based on the properties of nontrivial integration measures. 
We can understand the noncommutative geometry emerging as a consequence of the position uncertainty principle, and producing a kind of fuzzy space. This fuzziness region can be understood as a fractal property of spacetime, obviously, weakly realized at that range of energy.

\section{Concluding Remarks and Perspectives}

In summary, our main motivation is to develop an approach based on the fractional variational calculus. More precisely, the modified Riemann-Liouville fractional derivative, to address coarse-grained constrained systems, since FC shows itself a powerful tool to treat this class of systems.   Our effort is also to present a fractional symplectic algorithm for systems with a higher extent of complexity.  Consequently, we have shown that it is possible to discuss quantization in this scenario and clarify its correspondence by means of the  MRL approach.  We have  proposed two kinds of fractional formulations for the symplectic algorithm.  

The first formulation considers a Lagrangian model like $L=\eta f^{(0)}{}_0D^{\alpha}_t\eta-H(\eta)$  within the MRL approach.  We have obtained the Hamilton-Jacobi equations deformed by the fractional contribution in the symplectic scenario along with the symplectic matrix. The corresponding  Dirac brackets have also received the same kind of modification.

However, when we enlarge our class of applications by considering other  field theories; this first approach does not seem to be the more general way to treat them.  Therefore, we have convinced ourselves on  the necessity to generalize the Lagrangian by considering  $L=a_i(\eta){}_0D^{\alpha}_t\eta-V(\eta)$ . The constraints have been defined in the same way and the consequence was the extension of the symplectic matrix in a fractional framework.  We have obtained the final form for the fractional equations of motion and the symplectic matrix associated to the Dirac brackets, which  now exhibit an additional term $\Gamma(1+\alpha)$, due to the fractional contribution imposed by the fractional formalism.  An important point is that we can recover the usual brackets whenever $\alpha\rightarrow1$.

Evidently, other different definitions for the fractional derivative could be used with the same purpose.   For example, the generalized Euler formula, Abel or Fourier integral representation, Riemann-Liouville, Caputo, Sonin, Letnikov, Laurent, Nekrasov and Nishimoto representations, for example. But, we have to recall once more that, in all these approaches, the chain rule becomes rather messy, yielding very lengthy calculations. For this reason, perhaps, FC has found some resistance to become more familiar in  areas of Physics such as gravitation and field theory.

We also have illustrated our approach with an elucidative example where  a fractional Lagrangian extension was considered that describes charged particles in a magnetic field.  In this case, the time derivative has been changed by the fractional derivative, and the fractional simplectic algorithm was then implemented. We have obtained an $\alpha-$corrected expression for the Dirac brackets, and we have analyzed its approximated form whenever $\alpha\rightarrow0$. A very close analogy between $\alpha$ and the noncommutative parameter $\theta$ has been found out.

It is instructive to notice that the granularity of the system can be represented  by the fractional parameter $\alpha$ .  So, we can observe two different situations.  The first one, when $\alpha\rightarrow0$, we can think that the medium granularity  is gross, so the system can be understood as rough or sparse. In this limit, comparing the processes involved, the dynamics of the system can be understood as very slow, namely, the relaxations processes are slow, since the derivatives in the dynamical equations tend to zero. 

In our second limit, when $\alpha\rightarrow1^{-}$, we could talk about tiny or fine granularity, represented by the so-called low-level fractionality.  In this case, the system is almost continuous and not sparse. The interactions between parts of the systems and the environment are more frequent and the relaxations are faster.  The dynamics can be better understood and discussed in terms of complexity of these interactions. The anomalous behavior of the system can be connected with the non-locality of the interactions. Taking  this point of view, small discrepancies on the fundamental constants can be expected.

In the MRL approach, eq. (\ref{3}), we always work with $\alpha$ such that $0<\alpha<1$.  We could consider the extension of $\alpha$ to other ranges; specifically, if $\alpha<0$, we are in a regime of a different dynamics, whose equations are expressed in terms of integrals rather than derivatives. 

We have also come to the conclusion that a fractional model can provide us with a memory effect in the convolution integrals and leads to some differential equations which could open up different physical situations, such as viscoelasticity and more abstract scenarios such as mapping using tensorial fields.  

Besides the applications presented here, we strongly believe that quantization in a fractional context is a widely open area that may raise a tremendous deal of interest.  As a viable and almost immediate extension, we could consider the case of the coarse-grained formulation of a fractional Dirac equation without taking it as the $(\sqrt{\Box})$. We shall be reporting our studies on that in a forthcoming work.

We believe that a whole bunch of problems  can be handled using this approach.  Aspects of gravitation, condensed matter systems and field theory seem to be ready to be approached in terms of the FC and its tools.

\section{Acknowledgments}
J.A.Helay\"{e}l would like to express his gratitude to FAPERJ-RJ and CNPq-Brazil for the precious support to research projects.

C.F.L. Godinho would like to thank FAPERJ-RJ for financial support

C.F.L. Godinho and J. Weberszpil would like to thank CBPF-LAFEX  for the kind hospitality.


\begin{thebibliography} {99}
\bibitem{3}   P. A. M. Dirac, Lectures on Quantum Mechanics (Beffer Graduate School of Science, Yeshiva University, New York, 1964).
\bibitem{1}   M. Maeno "Canonical quantization of string fields: How to define momentum," Prog.Theor.Phys.Suppl. 188, 217-226 (2011); T. G. Philbin "Canonical quantization of macroscopic electromagnetism," New J.Phys. 12 123008 (2010).

\bibitem{fj} L.D. Faddeev and R. Jackiw "Hamiltonian Reduction of Unconstrained and Constrained Systems," Phys. Rev. Lett. 60,  1692-1694 (1988).


\bibitem{BNW1} J.Barcelos Neto and C. Wotzaseck "Symplectic quantization of constrained systems," Mod. Phys. Lett. A 7, 1737-1747 (1992). 


\bibitem{BNW2} J.Barcelos Neto and C. Wotzaseck "Faddeev-Jackiw quantization and constraints," Int. J. Mod. Phys. A 7, 4981-5003 (1992).


\bibitem{KS} K. Sundermeyer, "Constrained Dynamics: With Applications to Yang-Mills Theory, General Relativity, Classical Spin, Dual String Model : Lecture Notes in Physics" (Springer-Verlag, Berlin, New York, 1982). 


\bibitem{Shirzad} A. Shirzad and M. Mojiri "The Difficulty of symplectic analysis with second class systems," J.Math.Phys. 46, 012702 (2005).


\bibitem{RW}  F. Riewe "Nonconservative lagrangian and Hamiltonian mechanics" Phys. Rev. E 53, 1890-1899 (1996).

\bibitem{DY1} D.W. Dreisigmeyer and P.M Young "Nonconservative Lagrangian Mechanics: a generalized function approach" J. Phys. A: Math. Gen. 36 (2003) 8297-8310

\bibitem{DY2} D.W. Dreisigmeyer and P.M Young "Nonconservative Lagrangian Mechanics: a generalized function approach" J. Phys. A: Math. Gen. 37 (2004) L117-L121

\bibitem{Cal}   A. O. Caldeira and A. J. Legett "Influence of dissipation on quantum tunneling in macroscopic systems," Phys. Rev. Lett. 46, 211-214 (1981). 


\bibitem{Sen}   I. R. Senitzky "Dissipation in Quantum Mechanics. The Harmonic Oscillator" Phys. Rev. 119, 670-679 (1960).  

\bibitem{GZ}    G. Zaslavsky Hamiltonian Chaos and Fractional Dynamics, (Oxford University Press, New York, 2005).


\bibitem{referee sug 1}J.T. Machado, V. Kiryakova,F. Mainardi "Recent History of Fractional Calculus" Commun. Nonlinear Sci Numer. Simulat. 16 (2011) 1140–1153.


\bibitem{JTM book} J. Sabatier, O.P. Agrawal and J.A.T. Machado (editors) "Advances in Fractional Calculus:Theoretical Developments and Applications 
in Physics and Engineering" (Springer, Netherlands, 2007)


\bibitem{klafter} R. Metzler and J. Klafter "The random walk's guide to anomalous diffusion: a fractional dynamics approach" Physics Reports  339, 1-77 (2000). 


\bibitem{klafter2}R. Metzler and J. Klafter "The restaurant at the end of the random walk: recent developments
in fractional dynamics descriptions of anomalous dynamical processes" J. Phys. A: Math. Gen.  37, R161-R208 (2004). 


\bibitem{klafter3}R. Metzler, E. Barkai, and J. Klafter "Anomalous Diffusion and Relaxation Close to Thermal Equilibrium: A Fractional Fokker-Planck Equation Approach" Phys. Rev. Lett.  82, 3563-3567 (1999).


\bibitem{9}  M. A. E. Herzallah and D. Baleanu "Fractional Euler�Lagrange equations revisited" Nonlinear Dyn.  58, 385-391 (2009).

\bibitem{scalas} D. Fulger, E. Scalas, and G. Germano "Monte Carlo simulation of uncoupled continuous-time random walks yielding a stochastic solution of the space-time fractional diffusion equation"  Phys. Rev. E  77, 021122 (2008).


\bibitem{Hilfer2} R. Hilfer "Experimental evidence for fractional time evolution in glass forming materials" Chemical Physics 284, 399-408 (2002). 


\bibitem{stanisvasky} A. Stanislavsky1, K. Weron and J. Trzmiel "Subordination model of anomalous diffusion leading to the two-power-law relaxation responses" EPL  91, 40003 (2010).


\bibitem{Metzler} R. Metzler, W. Schick, H. G. Kilian and T. F. Nonnenmacher "Relaxation in filled polymers: A fractional calculus approach" J. Chem. Phys.  103, 7180-7186 (1995).


\bibitem{Glockle}  W. G. Glockle and T. F. Nonnenmacher "Fractional integral operators and Fox functions in the theory of viscoelasticity"   Macromol  24, 6426-6434 (1991).


\bibitem{cresus} Everton M.C Abreu and Cresus F.L. Godinho "Fractional Dirac Bracket and Quantization for Constrained Systems" Phys. Rev. E  84, 026608 (2011).


\bibitem{baleanu}D. Baleanu "About fractional quantization and fractional variational principles" Commun Nonlinear Sci Numer Simulat  14, 2520-2523 (2009). 


\bibitem{cal1} G. Calgagni "Geometry and field theory in multi-fractional spacetime" JHEP01 065 (2012).


\bibitem{goldfain} E. Goldfain "Derivation of lepton masses from
the chaotic regime of the linear $\sigma-model$" Chaos, Sol. and Frac.  14,  1331-1340 (2002).


\bibitem{goldfain2}E. Goldfain "Fractional dynamics, Cantorian space�time and the gauge hierarchy problem" Chaos, Sol. and Fractals  22, 513-520 (2004).


\bibitem{Herrmann} R. Herrmann "Gauge invariance in fractional field theories" Phys.Lett. A372, 5515-5522 (2008).


\bibitem{nos} C.F.L. Godinho, J.Weberszpil and J.A.Helay\"{e}l Neto "Extending the D'Alembert Solution to Space-Time Modified Riemann-Liouville Fractional Wave Equations" Chaos, Sol. and Fractals  45, 765-771 (2012).


\bibitem{nos2} J.Weberszpil,C.F.L. Godinho and J.A.Helay\"{e}l Neto "Aspects of the Coarse-Grained-Based Approach to a Low-Relativistic Fractional Schr\"{o}dinger Equation" PoS  027 ICMP (2012), arXiv:1206.2513.


\bibitem{referee sug 2}D. Val\'erio, J.J. Trujillo, M. Rivero, J. T. Machado and D. Baleanu "Fractional Calculus: A Survey of Usefull Formulas" Eur. Phys. J. Special Topics, 222, 1827-1846 (2013).


\bibitem{gj1} G. Jumarie "From Lagrangian mechanics fractal in space to space fractal Schr\"{o}dinger equation via fractional Taylor series" Chaos  Sol. and Frac. 41, 1590 (2009).


\bibitem{ric} R. Almeida, A. B. Malinowska and D. F. M. Torres "A fractional calculus of variations for multiple integrals with application to vibrating string" J. Math. Phys. 51 033503 (2010)

\bibitem{Kai} K. Diethelm, "The Analysis of Fractional Differential Equations: An Application-Oriented Exposition Using
Differential Operators of Caputo Type" Springer-Verlag Berlin Heidelberg (2010).

\bibitem{Go1} R. Gorenflo and F. Mainardi "Time-fractional derivatives in relaxation processes:  A tutorial survey" Fract. Calc. Appl. Anal. 10, 269-307 (2009)

\bibitem{Go2} R. Gorenflo and F. Mainardi, "Integral and differential equations of fractional order" A. Carpinteri and F. Mainardi (Editors),Fractals and Fractional Calculus in Continuum Mechanics, Springer Verlag, Wien and New York 1997, 223-276, ArXiv:0805.3823v1[math-ph]

\bibitem{lv} L. Longjin, F.Y. Ren and W.Y. Qiu "The application of fractional derivatives in stochastic models driven by fractional Brownian motion"  Phys. A 389, 4809-4818 (2010).


\bibitem{Khalili} A. K. Golmankhaneh et al "Hamiltonian Structure of Fractional First Order Lagrangian" Int. J. Theor. Phys. 49 Issue 2, 365-375 (2010). 


\bibitem{Zu}  W. H. Zurek "Decoherence, einselection, and the quantum origins of the classical" Rev. Mod.Phys. 73, 715-775 (2003).


\bibitem{Cv}  P. Cvitanovic, Universality in Chaos ($2^{nd}$ edition Adam
Hilger, Bristol, 1989).


\bibitem{Jackiw} R. Jackiw "Physical instances of noncommuting coordinates" Nucl Phys. Proc. Suppl. 108, 30-36 (2002).


\bibitem{Sus} L. Susskind "The Quantum Hall fluid and noncommutative Chern-Simons theory" arXiv:hep-th/0101029v3.


\bibitem{Michael Douglas} M. Douglas "Noncommutative field theory" Rev.Mod.Phys. 73, 977-1029 (2001).


\bibitem{Tara} V. E. Tarasov "Dynamics with low-level fractionality" Physica A  368, 399-415 (2006).



\bibitem{jecke}B. Jeckelmann and B. Jeanneret "The quantum Hall effect as an electrical resistance standard" Rep. Prog. Phys. 64, 1603-1655 (2001).

\bibitem{cal2} M. Arzano, G. Calcagni, D. Oriti and M. Scalisi "Fractional and noncommutative spacetimes" Phys. Rev. D 84, 125002 (2011)


\end{thebibliography}
\end{document}